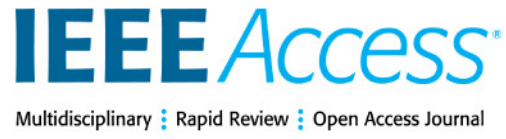

# To Code or Not to Code: When and How to Use Network Coding in Energy Harvesting Wireless Multi-hop Networks

**Nastooh Taheri javan, *(Senior Member, IEEE),* and Zahra Yaghoubi**

Computer Engineering Department, Imam Khomeini International University, Qazvin, IRAN.

Corresponding author: Nastooh Taheri Javan (e-mail: nastooh@eng.ikiu.ac.ir).

**ABSTRACT** The broadcast nature of communication in transmission media has driven the rise of network coding's popularity in wireless networks. Numerous benefits arise from employing network coding in multi-hop wireless networks, including enhanced throughput, reduced energy consumption, and decreased end-to-end delay. These advantages are a direct outcome of the minimized transmission count. This paper introduces a comprehensive framework to employ network coding in these networks. It refines decision-making at coding and decoding nodes simultaneously. The coding-nodes employ optimal stopping theory to find optimal moments for packet transmission. Meanwhile, the decoding-nodes dynamically decide, through SMDP (Semi Markov Decision Process) problem formulation, whether to conserve energy by deactivating radio units or to stay active for improved coding by overhearing packets. The proposed framework, named ENCODE, enables nodes to learn how and when to use network coding over time. Simulation results compare its performance with existing approaches. Our simulation results shed new light on when and how to use network coding in wireless multi-hop networks more effectively.

## I. INTRODUCTION

The theory of network coding remains relatively young as a research subject. Back in 2000, Professor Ahlswede and his team pioneered the concept of network coding for multicast applications in wired networks [1]. This approach empowers nodes to merge several input packets into one or multiple output packets, as opposed to merely relaying single packets. This innovation results in a reduction in the overall number of network transmissions [2].

Shortly after the introduction of network coding theory for multicast applications in wired networks, various studies investigations revealed that due to the nature of broadcast communication in wireless networks, network coding significantly contributes to enhancing network efficiency [3]. Notable benefits of network coding's integration into wireless networks encompass increased fault tolerance and packet recovery, improved throughput, and reduced energy consumption (due to the reduction in the number of transmissions) [4]. In this context, the term *coding gain* refers to the ratio of the required number of transmissions to send a specific number of packets without employing network coding, to the required number of transmissions for the same number of packets using network coding [5]; this value is always greater than or equal to one.

As an example of network coding implementation in wireless multi-hop networks, COPE can be mentioned [6]. In a nutshell, in COPE, each wireless node consistently listens to all transmitted packets from its neighboring nodes. Subsequently, each node is obligated to continually inform its neighbors about the list of packets stored in its memory. In this scenario, prior to sending a data packet, each node, based on the information received from its neighbors (including the list of packets stored in each neighbor's memory), engages in decision-making for coding and packet combination. This is done so that the node selects the best coding pattern, combines the packets together (using simple XOR), and sends the coded-packet. On the receiving end, nodes can recover their expected packets by XORing the received coded-packet with the packets they previously have stored in memory.

Reviewing previous research reveals that many existing implementations of network coding in multi-hop wireless networks have predominantly focused on maximizing the coding gain to improve the efficiency of network coding [7], often at the expense of other performance aspects [8]. However, we believe that by modifying certain decision-making processes at nodes, a rational balance between performance parameters while utilizing network coding can be achieved. The main goal of this study is to enhance the performance of network coding in multi-hop wireless networks through the refinement of decision-making procedures at nodes and providing an integrated framework for this purpose. To achieve this goal, two key approaches have been concurrently considered in this paper.

In the first approach, the emphasis is placed on the decision-making process at coding-nodes, while the second approach focuses on enhancing the decision-making process at decoding-nodes. By incorporating these two approaches collectively among network nodes, a proposed integrated framework takes shape. In this framework, nodes utilize the first approach to make decisions when engaged in data transmission. Conversely, they adopt the second decision-making process during different periods. Over time, nodes will learn when and how to effectively utilize network coding.

In the first decision-making approach, which we distinctively named *To-Send-or-Not-to-Send* in one of our previous research studies [9], nodes during the encoding process strive to make the best decisions for selecting the coding degree of the packets intended for transmission. Here, the term *coding degree* refers to the number of original packets that are combined together in a coded packet [10]. In this scenario, when a coding-node makes a decision to transmit a packet, it must select the optimal coding pattern for it. If the coding-node intentionally delays the transmission of a packet, there is a possibility that better coding patterns could emerge based on the arrival of reports on existing packets stored in neighboring nodes' memories. However, deliberate postponement of transmissions implies intentional increase in end-to-end delay within the network. In this approach, nodes employ optimal stopping theory [11] to model the problem. Essentially, nodes engage in a trade-off between enhancing coding gain and reducing delay in the network. Over time, nodes learn the optimal timing for employing network coding and transmitting packets. The question arises: What is the best time for nodes to transmit packets and use network coding?

In the second decision-making approach, which we separately termed in one of our previous studies *To-Overhear-or-Not-to-Overhear* [12], network nodes aim to gradually learn the best decision between staying awake (and listening to new packets) or going to sleep (and conserving energy). One of the significant issues in employing network coding in wireless networks is that the more a node overhears packets from others, stores these packets in its memory, and reports them to its neighbors, the more it contributes to enhancing coding efficiency in neighboring nodes. This improvement in coding efficiency in wireless networks is so significant that in most implementations, all nodes during their idle time are engaged in overhearing packets being transmitted among their neighbors; this behavior is termed *Overhearing* [13]. In this approach outlined in our research, nodes strive to discover an appropriate pattern for their sleep/wake-up over time during idle periods based on coding efficiency. Using a model based on semi-Markov decision processes [14], nodes in this approach make a trade-off between improving coding gain and reducing energy consumption and finally learn how to use network coding effectively.

In the current paper, both aforementioned decision-making approaches have been simultaneously implemented within network nodes, providing an integrated framework for using network coding in energy harvesting multi-hop wireless networks. To achieve this, within the proposed framework titles ENCODE[1], certain modifications have been applied to the system model. The majority of these modifications are centered around enhancement of the IEEE 802.11 protocol [15] for the MAC layer and refining the node energy model in harvesting process from the environment. Subsequently, the impact of each decision-making approach on the performance of the other is elucidated through simulation. Following that, both decision-making processes are implemented concurrently within a proposed integrated framework. Their performance is then compared against both the COPE reference approach and a scenario without the using of network coding. It is noteworthy that this study considers four main evaluation metrics for assessing the efficiency of network coding: coding gain, end-to-end delay, throughput, and energy consumption.

The main contributions of this paper can be summarized as follows:

- Evaluation of the impact of the two mentioned decision-making approaches on each other's performance individually.
- Revising the to-overhear-or-not-to-overhear approach through enhancing the procedure of evaluating the usefulness of the overheard packets.
- Enhancing the System model to facilitate the simultaneous implementation of both decision-making approaches in network nodes.
- Presenting an integrated framework for using network coding in energy harvesting wireless multi-hop networks, named ENCODE, and comparing its performance with existing approaches.

The following sections of this paper are organized as follows. In the second chapter, related research will be examined. The third chapter elaborates on the main motivation of this paper, while the fourth chapter discusses the system model. In the fifth chapter, the proposed framework is described, followed by the presentation of numerical results obtained from simulations in the sixth chapter. Finally, the seventh chapter provides a summary and conclusion of this study.

[1] Efficient Network CODing.

## II. Related Works

In this section, first, the details of COPE as the reference model have been examined, and subsequently, the research related to both decision-making approaches has been reviewed separately.

### A. COPE

One of the most impactful implementations in the field of inter-flow network coding applications in wireless networks is an idea known as COPE [6]. In the COPE architecture, an encoding layer is added between the IP and MAC layers. The primary task of this layer is to examine and identify coding opportunities, combine two or more packets into a coded packet, and transmit it to neighbors. The COPE concept is essentially composed of three main phases.

**Opportunistic overhearing:** In the COPE architecture, due to the nature of broadcast transmission medium, all nodes remain active in promiscuous mode. As a result, each node overhears all transmitted packets in its vicinity and temporarily stores them in memory. Furthermore, each node is obligated to share the statistics of the packets stored in its memory by sending reception reports to its neighboring nodes.

**Opportunistic Coding:** One of the most significant questions raised in the realm of network coding is: *Which packets should be coded together?* When making decisions for this purpose, a node may have several options and choices. The primary concern in this selection process is that the encoding node attempts to include the maximum number of packets in a coded packet, provided that the receiving nodes have the capability to recover the original packets from the coded packet.

**Neighbor State Learning:** Each node needs to be aware of the buffer content of its neighboring nodes. For this reason, nodes send reception reports to their neighbors. On the other hand, after receiving reception reports from their neighbors, nodes can make better decisions for packet encoding in the coding phase.

### B. Related Research on Decision-Making in Coding-Nodes

Some recent studies in the field of using network coding in wireless networks have been focused on the question of whether packets that lack suitable coding opportunities in intermediate nodes can be intentionally delayed or not. However, the majority of these investigations have imposed highly restrictive assumptions for addressing this issue. The most crucial limiting assumption noticeable in these studies is the adoption of the *Reverse Carpooling* scenario (also known as two-way-relay scenario) [16]. In this scenario, two specific flows travel in precisely opposite directions within the target intermediate node, and this node aims to encode and transmit packets from these two flows together.

In [17], the issue of retaining packets in coding-nodes has been modeled using Markov chains and hidden Markov models in the Reverse Carpooling scenario. In [18], the same scenario has been formulated as a Markov decision process, and a solution has been provided using stochastic dynamic programming. In [19], authors have proposed an opportunistic scheduling mechanism for the Reverse Carpooling scenario, where nodes decide based on their buffer state on one side and the channel state on the other side. In [20], the decision-making problem concerning increasing packet delay to enhance coding opportunities has been examined specifically in the context of video transmission and the Reverse Carpooling scenario. In [21], a solution has been presented for the Reverse Carpooling scenario using the Primal-dual method to minimize the overall system cost, including delay and the number of transmissions.

In [22], authors have modeled the Reverse Carpooling scenario as a continuous-time Markov chain and demonstrated that these networks are exactly correspondent to the positive-negative customer problem. In [23], the authors focus on the cost of transmissions and delays in the Reverse Carpooling scenario, and this scenario is modeled as a Markov decision process. In [24], authors model the energy-delay trade-off in the Reverse Carpooling scenario using time-wait policies. In [25], authors model the Reverse Carpooling scenario using a discrete-time Markov chain and model the packet arrival process using a discrete-time Markovian arrival process. In [26], authors propose a distributed strategy based on game theory to optimize the energy-delay trade-off in the Reverse Carpooling scenario. In [27], authors present a frequency division multiplexing technique for the Reverse Carpooling scenario. In [28], the problem is confined to a limited train network, and a cross-layer strategy between the network layer and the data link layer is devised.

### C. Related Research on Decision-Making in Decoding-Nodes

Some studies attempt to reduce the level of overhearing in limited scenarios for network-coded wireless communication networks. Some achieve this by restricting the volume of received packets, while others employ sleep/wake-up mechanisms.

In [29], a solution is proposed to reduce redundant packet overhearing in wireless sensor networks for multicast applications. In this method, each node broadcasts a small packet titled *digest info* before transmitting the main packet. This digest info packet contains a summary of the upcoming main packet's data. Other nodes can use this small packet to determine the repetitiveness or novelty of the subsequent main data packet. Based on this information, nodes can make decisions for their sleep/wake-up patterns. In [30], network coding is considered in GinMAC networks. To decrease the overhead of packet overhearing, the authors attempt to mitigate some of the overhearing time-slots by dynamically re-allocating time-slots (altering, removing, or adding certain time slots). This adjustment aims to reduce the number of overhearing time-slots and enhance network efficiency.

In [31], the authors have attempted to provide an integrated solution of network coding and sleep/wake-up scheduling for bottleneck points in wireless sensor networks with multimedia applications, using energy distribution among neighboring nodes of the sink node. In [32], a combined framework is proposed for wireless sensor networks with flooding applications, aiming to merge network coding with sleep/wake-up scheduling to address the challenge of redundant transmissions in multicast-based solutions. In [33], the authors introduce a solution called GreenCode, essentially a coding-aware cooperative channel access protocol, which aims to enhance energy efficiency through sleep/wake-up scheduling of nodes.

In some recent studies, the decision-making process at destination nodes for decoding operations is communicated to the source node in the form of feedback [34]. Feedback-Based Network-Coded Systems combine the benefits of network coding with real-time feedback to create adaptive and efficient communication systems in dynamic and challenging network environments. The integration of feedback allows nodes to make informed decisions, improving reliability and optimizing throughput [35].

## III. Preliminaries and Motivation

### A. Preliminaries

The utilization of network coding in wireless networks, due to its reduction in the number of transmissions, offers diverse advantages such as energy consumption reduction and enhancement of network throughput. Many prominent research efforts in this field have predominantly focused on enhancing the coding gain. In our prior two studies, we also concentrated on two primary trade-offs. In [9], addressing the to-send-or-not-to-send problem, we examined a trade-off between enhancing coding gain on one hand and reducing end-to-end delay on the other. Similarly, in [12], tackling the to-overhear-or-not-to-overhear problem, we investigated a trade-off between increasing coding gain and diminishing energy consumption in multi-hop wireless networks.

In the first issue, we examined the problem of delaying the transmission of coded-packets in a general scenario, without considering the reverse carpooling assumption, with the aim of discovering a more suitable coding pattern. As mentioned in section 2-B, in some prior studies such as [17-19], the reverse carpooling scenario has been investigated only for two similar flows with opposite directions. This scenario and its associated assumptions are highly restricted and specific. However, we have addressed this problem in a general model without any limiting assumptions. In the second issue, for the first time, we formulated the problem of avoiding the overhearing of packets that do not provide any coding gain for the purpose of improving coding efficiency. We employed semi-Markov Decision Processes (SMDP) and applied a reinforcement learning solution [36] for modeling.

The results obtained from the separate implementation of these two decision-making approaches were very promising. The first approach led to significant enhancements in terms of network throughput and energy consumption. Meanwhile, the implementation of the second approach resulted in a noticeable increase in the network's lifetime compared to existing methods like COPE [6], which cannot be disregarded.

### B. Motivation

Continuing these two research endeavors, we decided to simultaneously implement both approaches at the nodes and propose an integrated framework named ENCODE. In this framework, nodes employ the first approach during packet transmission, while utilizing the second approach for decision-making during other time intervals. This cohesive strategy aims to synergize the benefits of both approaches and contribute to a more efficient and resilient wireless network.

The overall flow of the ENCODE framework is well illustrated in Fig. 1, depicting it across four consecutive time slots. In this figure, the time instances labeled as $t_i$, $t_j$, $t_k$, and $t_l$ are sequentially examined. In this scenario, we assume the existence of two flows: one from node $n1$ to node $n4$ and another from node $n4$ to node $n2$. These flows are established in a way that, based on the considered topology, they converge at node $R$.

In Fig. 1-a, node $n1$ has transmitted packet $a$ towards node $R$, providing an opportunity for both nodes $n2$ and $n3$ to overhear this packet. Utilizing the second approach within the ENCODE framework, node $n2$ considers this packet $a$ as valuable for its neighboring nodes' coding operations and overhears it. However, node $n3$ prefers to go to sleep instead of overhearing packet $a$ during its transmission.

In Fig. 1-b, at time moment $t_j$, while packet $b$ has arrived from node $n4$ to node $R$, a transmission opportunity arises for node $R$. At this instant, node $R$ is unaware of the overhearing of packet $a$ by node $n2$ and consequently is not aware of the available coding opportunity. Following the to-send-or-not-to-send approach, at this moment, node $R$ prefers to delay the transmission process slightly in the hopes of obtaining better coding opportunities.

In Fig. 1-c, at time instant $t_k$, node $n2$ sends its reception report to its neighbors, including node $R$. After receiving this report, node $R$ discovers a suitable coding opportunity for combining packets $a$ and $b$. As a result, during its next transmission opportunity at time $t_l$, node $R$ sends the coded-packet $a+b$ to nodes $n2$ and $n4$ in a single transmission.

Indeed, the decision to postpone the transmission for node $R$ at time $t_j$ based on the to-send-or-not-to-end approach, coupled with node $n2$ overhearing on packet $a$ while node $n3$ abstained from overhearing on the same packet following the to-overhear-or-not-to-overhear approach, all represent rational and sensible actions. These decisions collectively demonstrate a strategic balance between optimizing coding opportunities and energy-efficient behavior within the network.

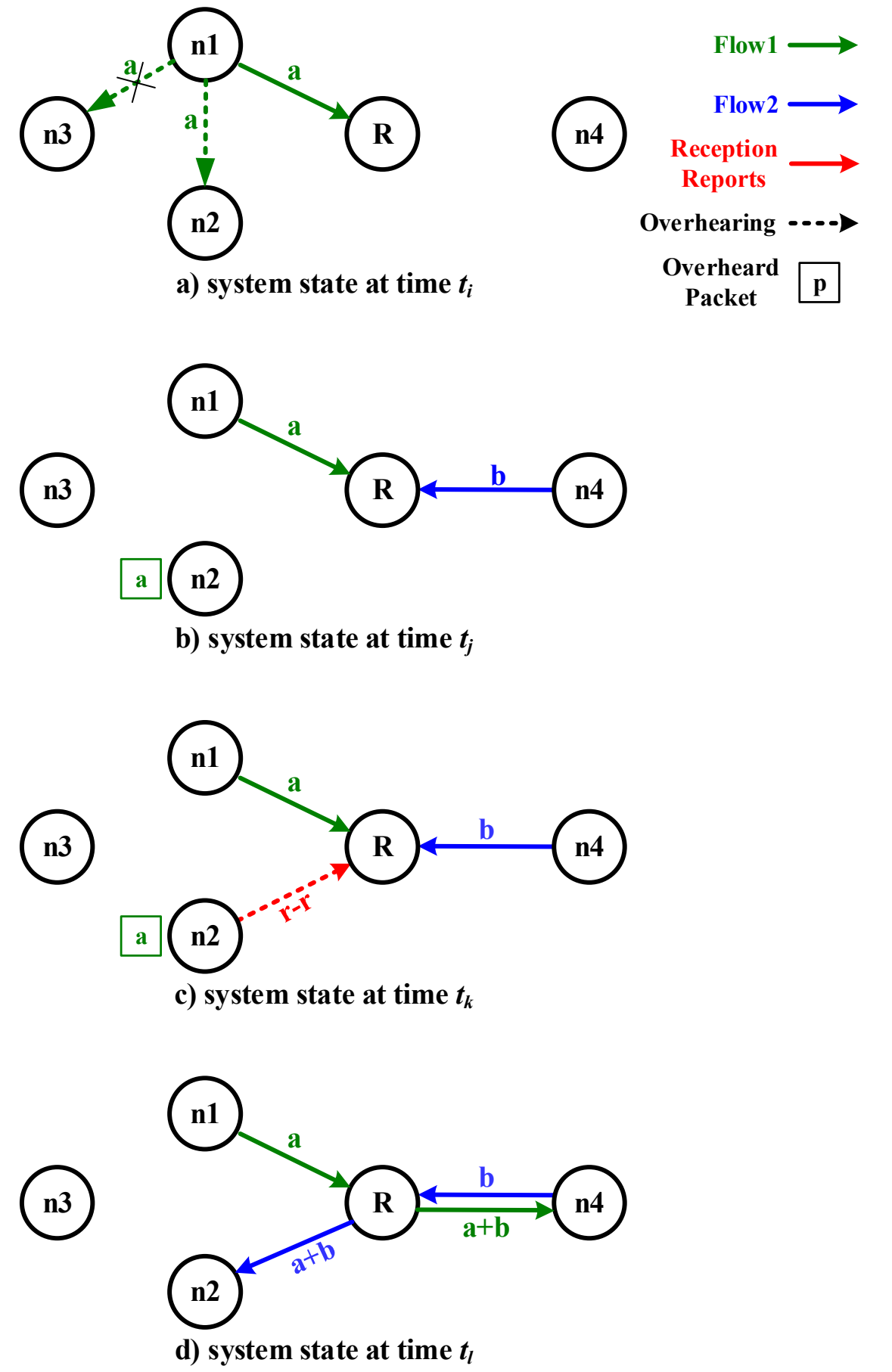

**FIGURE 1.** A simple example of how ENCODE works.

The scenario depicted in Fig. 1 has progressed optimistically, but a crucial detail that emerged during the practical implementation of the ENCODE framework has been overlooked. After the simultaneous implementation of two decision-making approaches, we realized that these two approaches jointly influence performance. One of the most significant impacts is that under certain conditions, Reception Reports may not reach neighboring nodes or may be delivered to them with delay. Therefore, the concurrent implementation of these two decision-making approaches and the examination of the obtained results have been the primary motivation for writing this article.

Consider the following scenario as an example: Fig. 1-c and the moment $t_k$ need to be meticulously examined. The critical issue is that at time $t_k$, when node $n2$ sends the reception report packet to its neighbors, node $R$ might perceive the data within this packet as non-beneficial for coding operations and consequently, it could disregard overhearing this packet. As a result, it might fail to recognize the coding opportunity that has emerged. Consequently, even during the subsequent transmission instance, node $R$ still lacks a suitable coding opportunity for the packets it is currently sending.

The mentioned issue arises from the fact that in the to-overhear-or-not-to-overhear approach, the value of packets for overhearing was solely determined based on their data content. However, it seems that the reception reports present within the packets should also have a significant influence on their value for overhearing. To address this problem, within the ENCODE framework, we have modified how the value of packets for overhearing is evaluated. In this manner, not only the data contained within the packets but also the reception reports present in the packets affect the value of the packets for overhearing.

## IV. System Model

In this section, the system model and key assumptions are presented separately for the network model, coding model, energy model, and MAC model.

### *A. Network Model*

In this study, a multi-hop wireless network with stationary nodes is considered. In this network, nodes are randomly distributed in the environment, where each node can serve as the source or destination of traffic, and half-duplex connections are established between them. The network's topology is described by a graph $G = (N, H)$, where $N$ is the set of nodes and $H$ is the set of directed connections between nodes. Each node can establish connections with its neighbors within a radius of $\rho$.

When establishing a flow, initially, a node is selected as the source and another node as the destination of the flow. Subsequently, a stream of packets is exchanged between these two nodes. We have assumed, similar to many network coding implementations [6], that all transmitted data packets are of the same size. In cases where transmitted packets have different sizes, the source (or coding-node) appends extra bits (with a value of zero) to the end of smaller packets as needed.

In the entire network, we have assumed a uniform node architecture. For example, all nodes utilize omni-directional antennas and operate in a half-duplex mode. Network nodes can listen to their one-hop neighbors in a promiscuous manner. Nodes maintain a queue for outgoing packets to their neighbors, denoted by the parameter $p$ indicating the number of packets in this queue. Apart from this queue, nodes maintain another queue for storing packets they have overheard from their neighbors, represented by a size of $q$ packets. The overheard packets are stored in memory for a limited duration and will be sequentially discarded after a certain period, aided by an *aging* algorithm.

### *B. Coding Model*

In this study, we have assumed that a coding node is a node that combines various packets (from different flows) together using a specific coding scheme (here, XOR). Destination nodes that receive these coded-packets can utilize other packets stored in their memory to decode the received coded-packet and retrieve their intended packet. These nodes are referred to as decoding nodes.

In some studies, network coding is executed through linear combinations over a finite field, also known as a Galois Field (GF). Using a large Galois field presents a drawback due to high computational costs for packet encoding and decoding. Unlike GF (2) operations that rely on XOR, large Galois fields involve multiplication and Gaussian elimination. This complexity results in impractical energy consumption, particularly for battery-constrained devices like mobile phones and wireless sensors, especially when tackling coding-decoding challenges.

In this approach, for selecting the best possible coding option, the encoding node needs to possess a list of packets available to its neighbors. To gather this information, each node in the network sends a comprehensive list of packets stored in its memory to its neighbors, labeled as reception reports. However, in practice, there is no need to send separate packets for transmitting these reception reports; instead, reception reports are transmitted alongside regular data packets. Each data packet can carry multiple reception reports, where each report includes the overheard packets and the sender from whom those packets were received. To perform coding operations, a specific header is added to data packets, placed between the network layer header and the data link layer header. This header is referred to as the coding header. The coding header comprises three main components: 1) Identification of packets participating in the coding of the current packet, 2) Reception reports, and 3) Acknowledgment of received packets.

Each data packet can carry multiple Reception Reports, each consisting of received packets and the sender from which the packets were received. For this purpose, in the reports, the source address of the packets is first recorded, followed by the identifier of the last received packet from that source, and then a bit sequence for recently received packets from that source is registered. For example, the format of a report may be as follows: {23.115.61.202; 231; 001010010010011} This report signifies that the sender node of this report has several new packets in its memory, with the last received packet being packet 231. Specifically, packets with identifiers 231, 230, 227, 224, 221, and 219 have been received from that node. It is evident that in this method, the reception (or non-reception) of a specific packet may be reported multiple times, along with several data packets. For instance, the reception of packet 230 may also be mentioned in the next report.

On the other hand, when a node receives a coded-packet, it consults the coding header of that packet. This coding header contains a list of packet identifications that were involved in the coding process. Let's assume that the coding node of this coded-packet XORed $n$ original packets together and transmitted it as the current coded-packet for this receiving node. Now, this node needs to recover the mentioned original packets from its memory. According to the decoding procedure, this node should have $n$-1 packets among the packets that were coded together. By XORing these packets with the received coded-packet, the node can retrieve its expected packet.

### C. Energy model

In this paper, it is assumed that each node supplies its energy through a rechargeable battery. Additionally, the node attempts to utilize environmental energy sources during its lifetime. However, once the battery reserves are depleted, the node will be turned off. Recently, significant research advancements have been made in the field of harvesting energy from environmental sources, and efficient methods for harvesting energy from sources such as solar radiation, vibration, and ambient heat have been proposed for wireless nodes as well [37]. In this study, it is presumed that nodes accumulate the harvested energy in rechargeable batteries for future use. Naturally, the energy consumption rate doesn't exactly match the energy harvesting rate in practice, and we have assumed that the timing of energy arrivals to a node follows a stochastic process [38]. Fig. 2 illustrates the proposed model for energy harvesting and consumption in nodes, where $e_t$ represents the amount of energy available in the battery at time $t$.

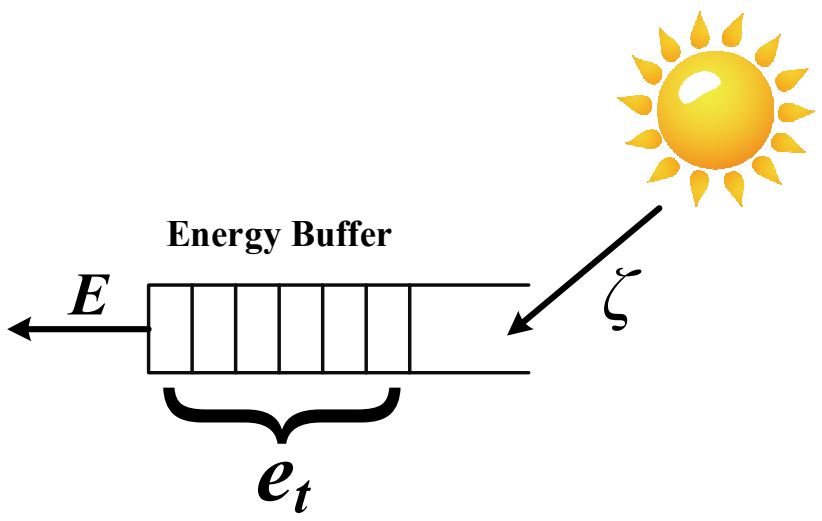

**FIGURE 2. Energy harvesting model.**

Furthermore, we have assumed that each node consumes an amount of $E_T$ energy units during the transmission of a packet. Similarly, for receiving (or overhearing) a packet, $E_R$ energy units are required. Finally, within a time interval equal to the transmission of a packet, if a node doesn't receive any packets (in an idle listening state), it consumes $E_I$ energy units. In the case where a node goes into a sleep mode, we have assumed that the energy consumption can be neglected.

### D. MAC model

In this study, the nodes at the MAC layer employ the IEEE 802.11 standard protocol [39], which utilizes CSMA/CA for channel access. The 802.11 standard has two modes: unicast transmission and broadcast transmission. In unicast transmission mode, the receiver immediately acknowledges the receipt of each packet, and the absence of an acknowledgment within a specified time interval indicates a collision, prompting the sender to retransmit the packet. This enhances reliability. However, in broadcast transmission mode, where a packet has multiple different destinations, acknowledgments from receivers are disregarded. In this study, we employ the unicast transmission mode of the 802.11 standard, enabling us to take advantage of its reliability benefits.

It has been demonstrated that the inter-arrival time between two consecutive transmission opportunities, denoted as $T$, for a specific node follows an exponential distribution with parameter $\lambda_t$. In practice, the availability of transmission opportunities for a particular node is dependent on factors and parameters such as network topology, congestion, neighboring nodes' traffic patterns, and other related parameters. Thus, the timing of transmission opportunities' occurrence for a node can be considered a random variable. We assume that packets arrive at a node by a Poisson distribution with parameter $\lambda_p$. Furthermore, the packet generation process at nodes follows i.i.d. Poisson processes. Consequently, it can be stated that the number of received packets by a node in a time interval $t$ follows a Poisson distribution with parameter $\lambda_p.t$.

## V. ENCODE: The Proposed Integrated Framework

In this section, the details of ENCODE are elaborated upon. Initially, the decision-making process in the coding nodes is summarized, similar to the to-send-or-not-to-send approach. Following that, by revising the old to-overhear-or-not-to-overhear approach, a new decision-making process for the decoding nodes is presented. Finally, some necessary modifications to the IEEE 802.11 protocol for implementing the proposed framework are also outlined.

### A. Decision-making approach in coding nodes

The inherent randomness of packet arrivals at the coding nodes poses a distinct challenge in network coding theory. Consider a coding node with a packet in its transmission queue intended for destination $X$, and it is now its turn to transmit this packet. Leveraging the concept of network coding, the sender node searches its transmission queue for coding opportunities and examines the other packets to identify a suitable coding pattern. However, let's assume that this node does not find any coding opportunity for the packet currently being transmitted! In this scenario, the node faces a critical decision and two choices:

- **First choice:** Transmit the intended packet purely to the destination without any manipulation. By choosing this option, the node essentially disregards the advantages of network coding and promptly transmits the packet to minimize end-to-end delay.
- **Second choice:** Retain the intended packet for a while in the hope of encountering coding opportunities.

The coding-node is aware that its neighbors are actively listening to various packets and continuously transmitting reports of the received and overheard packets to it. Therefore, there is a prospect that amidst the received packet reports, a suitable coding opportunity for this packet might arise. If such a suitable coding opportunity is found, the node can leverage the advantages of network coding to enhance network efficiency.

However, the main question is: if the second option is chosen (i. e. waiting for better coding opportunities), how long should the node wait to discover a coding opportunity? The crucial issue here is the incurred delay on the mentioned packet. Indeed, there's a possibility that excessive delay in retaining the packet, coupled with increasing end-to-end delay, might lead to the packet being perceived as lost from the perspective of upper layers (such as TCP or even the application).

We have addressed this decision-making in coding nodes under the title of the to-send-or-not-to-send in our previous study [9], which we briefly review in this section. Considering the nature of the decision-making conditions of this problem, we have modeled it using optimal stopping theory. Optimal stopping-based problems generally involve situations in which an agent needs to determine the best time to stop (or perform an action). In our problem, a sequential decision-making process is established for each node, where the node must choose between two options for its next action: continuing to wait or transmitting the packet.

In this problem, over time, the node receives reports of received packets from its neighboring nodes. Each of these reports holds a specific value in the context of enhancing coding gain. These reports and their corresponding values are essentially observable random variables for the learning agent in the optimal stopping theory. The learning node, by observing these variables, needs to determine the appropriate timing for issuing the stop command (packet transmission). The nature of the current problem belongs to the second type of optimal stopping problems, meaning it allows for recalling past observations.

#### 1) FORMULATION

In optimal stopping theory, the main problem is to choose a time for performing a specific action by observing a sequence of consecutive random variables, in a way that either maximizes the payoff or minimizes the cost. In this theory, a sequence of observable random variables is available to the learning agent. Depending on the problem type (finite horizon or infinite horizon), The agent can choose to receive and observe this sequence for as long as desired, but after observing one of these values, it must issue a stop command and receive a specific reward.

Our problem was modeled using a quadruple $\langle D, A, R, P \rangle$, where $D$ represents the state space, $A$ denotes the set of actions, $R$ stands for the reward function, and $P$ represents the transition probability matrix.

**Stage:** In this problem, each stage corresponds to the time interval between two consecutive transmission opportunities. Therefore, in each available transmission opportunity, the node decides whether to transmit or wait for better coding opportunities.

**State variable:** The state of a node, denoted as $d$, corresponds to the highest coding degree for all packets in the transmission queue of a node. In this case, the set $D = \{1, 2, 3, ...\}$ encompasses all possible states.

**Action set:** In each decision epoch, a node has two options to choose from: selecting 0 for waiting and selecting 1 for transmission (stop). Therefore, the action set $A$ is equal to {0, 1}.

**Transition matrix:** denoted as $P$, is defined based on the assumption that $p_{ij}(a, T_0)$ represents the probability of transitioning from state $i$ to state $j$ when action $a$ is chosen and the next stage occurs after a time interval of $T_0$. Accordingly, the transition matrix is defined as follows:

$$P = [p_{ij}(a, T_0)] \quad (1)$$

Now by taking $D$ to represent the set of all possible states we have:

$$p_{ij}(1, T_0) = p\{x_{t+1} = j | x_t = i, a = 1, T = T_0\}; i, j \epsilon D \quad (2)$$

$$p_{ij}(1, T_0) = 0; \ \forall j \epsilon D - \{1\} \quad (3)$$

$$p_{i1}(1, T_0) = 1; \qquad \forall i \epsilon D \quad (4)$$

The equations presented above suggest that opting for action 1 (transmitting the encoded packets) results in the system transitioning to state 1 (where $d = 1$) and remains in that state. Notably, State 1 functions as an absorbing state. Additionally, the following relationship holds:

$$p_{ij}(0, T_0) = p\{x_{t+1} = j | x_t = i, a = 0, T = T_0\} = \begin{cases} 0 & ; j < i \\ \dfrac{e^{-\lambda_d T_0} (\lambda_d T_0)^{j-i}}{(j-i)!} & ; j \geq i \end{cases} \quad (5)$$

This indicates that selecting action 0 (waiting for another epoch) in each iteration steers the system towards a more favorable condition, specifically one with a higher value of $d$. In this state, the decision to wait and receive packets does not lead to a decrease in $d$.

**Reward function:** $R(d, a)$ represents the obtained reward when taking action $a$ in state $s$. The function $R$ is defined as follows:

$$R(d, a) = \begin{cases} g(d-1) & ; a = 1 \\ 0 & ; a = 0 \end{cases} \quad (6)$$

where $g(.)$ represents the benefit gained from reducing the number of transmissions, such that when a coded packet with degree $d$ is transmitted, exactly $d$-1 transmissions are saved in the network. On the other hand, when the packet transmission is delayed until the next decision epoch, the system incurs a cost associated with the delay of packets. The achievable reward sequence for a node can be formulated as follows:

$$Y_0 = 0$$
$$Y_n(d_1, d_2, \dots, d_n, T_n) = g(d_n - 1) \times e^{-L.\delta.T_n}; n \geq 1 \quad (7)$$

where $T_n$ represents the incurred delay, $L$ is the maximum number of packets in the transmission queue (or the buffer size), and $\delta$ is a discount factor. In the above expression, the coding benefit decreases based on the incurred delay.

As a matter of fact, the achieved coding gain is discounted by the delay experimented in coding. We use the exponential discount of the reward because exponential discount is monotone and can monotonically decrease the reward. This structure is helpful for developing control policies. Moreover, the exponential discount factor can also handle the additive delay.

### 2) STOPPING RULE AND OPTIMAL SOLUTION

In our problem, to find the optimal control policy, we need to use the 1-SLA[2] approach to determine a threshold on the node's coding degree. In the 1-SLA method, at each stage, the expected achievable reward for one future stage is calculated. Then, the obtained reward at the current stage is compared with the expected reward of the next stage. If the current stage's reward is greater, the observations are stopped, and the decision to stop is made. Otherwise, the decision to continue is made.

In [40], it has been proven that for problems where the conditions $E\{\sup_n Y_n\} < \infty$ and $\lim \sup_{n\to\infty} Y_n \leq Y_\infty$ hold, then the 1-SLA stopping rule is optimal. Moreover, in [9], it has been demonstrated that these two conditions hold in our problem. Finally, the optimality equation in the current problem is written as follows:

$$v(d) = max\left\{\int_0^\infty \sum_{j\geq d} p_{dj}(0, T) \times v(j) \times e^{-L.\delta.T} f_T(t) dt, g(d-1)\right\} \quad (8)$$

where $v(d)$ is equal to the maximum expected achievable reward when the system is in state $d$.

In real-world scenarios, the coding degree of packets in the transmission queue is influenced by two primary factors: 1) the reception of new data packets, and 2) the receipt of reception reports. Both of these factors have the potential to create fresh coding opportunities for the existing packets in a node's transmission queue. We make the assumption that packets enter nodes following a Poisson distribution with a parameter of $\lambda_p$. Additionally, we assume that Reception Reports are received based on a Poisson distribution with a parameter of $\lambda_r$.

Under these assumptions, each newly received packet has the probability $p_p$ of incrementing the best coding level of existing packets (i.e., $d$), and each new Reception Report also has the probability $p_r$ of incrementing the best coding level of existing packets. Taking these considerations into account, the overall rate of increment for the degree of the best coding opportunity, denoted as $\lambda_d$, is determined as follows:

$$\lambda_d = \lambda_r . p_r + \lambda_p . p_p \quad (9)$$

To obtain the optimal stopping solution, we need to compare the reward of stopping in the current stage with the expected reward of stopping in the next stage, in order to determine when waiting is advantageous and under what

[2] 1-Stage Look Ahead.

conditions waiting for more reward is not worthwhile. To achieve this goal, $B$ is defined as follows:

$$B = \left\{ d: g(d-1) \geq \int_0^\infty \sum_{j \geq d} p_{dj}(0,T) \times g(j-1) \times e^{-L.\delta.T} f_T(t)dt \right\} \quad (10)$$

where $B$ represents the set of all states which stopping in that state is at least as beneficial as continuing and stopping in the next stage. After some simplification, we will have:

$$B = \left\{ d: g(d-1) \geq \int_0^\infty \sum_{j \geq d} \frac{e^{-\lambda_d T}(\lambda_d T)^{j-d}}{(j-d)!} \times g(j-1) \times e^{-L.\delta.T} \times \lambda_t e^{-\lambda_t T} dT \right\} \quad (11)$$

Next, it needs to be demonstrated that the set $B$ is a closed set over $d$, and subsequently, using the 1-SLA approach, we can deduce an optimal stopping rule. If we assume that sending a packet with coding degree $d$ can save us from $d$-1 transmissions in the network, then $g(.)$ can be linearly approximated as $g(d) = cd + b$. Consequently, the set $B$ will be defined as follows:

$$B = \left\{ d: cd + b \geq \int_0^\infty \sum_{j \geq d} \frac{e^{-\lambda_d T}(\lambda_d T)^{j-d}}{(j-d)!} \times (cj+b) \times e^{-L.\delta.T} \times \lambda_t e^{-\lambda_t T} dT \right\}$$

$$= \left\{ d: cd + b \geq cd.E_T\left(e^{-L.\delta.T}\right) + c\lambda_d E_T\left(Te^{-L.\delta.T}\right) + b.E_T\left(e^{-L.\delta.T}\right) \right\}$$

$$= \left\{ d: \left(1 - E_T\left(e^{-L.\delta.T}\right)\right)(cd+b) \geq c\lambda_d E_T\left(Te^{-L.\delta.T}\right) \right\} \quad (12)$$

$$= \left\{ d: d \geq \frac{\lambda_d E_T\left(Te^{-L.\delta.T}\right)}{\left(1 - E_T(e^{-L.\delta.T})\right)} - \frac{b}{c} \right\}$$

$$= \left\{ d: d \geq \frac{\lambda_d \frac{\lambda_t}{(\delta L + \lambda_t)^2}}{\left(1 - \frac{\lambda_t}{\delta L + \lambda_t}\right)} - \frac{b}{c} \right\}$$

$$= \left\{ d: d \geq \lambda_d \frac{\lambda_t}{\delta L(\delta L + \lambda_t)} - \frac{b}{c} \right\}$$

We have proved in [9] that the stopping rule (11) is an optimal solution for our problem. Thus, the obtained threshold for the decision-making process in our problem is given by:

$$d^* = \lambda_d \frac{\lambda_t}{\delta L(\delta L + \lambda_t)} - \frac{b}{c} \quad (13)$$

And finally, $w(d)$, the decision-making rule in this problem is given by:

$$w(d) = \begin{cases} 0; & d < d^* \\ 1; & d \geq d^* \end{cases} \quad (14)$$

This means that in each decision epoch, each coding node must first calculate the value of $d^*$ and compare it with the current value of $d$. If the current $d$ is greater than $d^*$, the node decides to stop and transmit the packet in that epoch. Otherwise, the node must wait until the next decision epoch.

In practice, the value of $d^*$ is influenced by two main factors: 1) the rate of occurrence of coding opportunities and 2) the rate of occurrence of transmission opportunities. If the rate of coding opportunities is high, it leads to a higher threshold value. This means that it's logical for a node to wait more in this situation, hoping for better coding degrees. On the other hand, if transmission opportunities are rare, the threshold value is calculated to be lower. This indicates that due to the scarcity of transmission chances, the coding node might be satisfied with transmitting packets with lower coding degrees.

### B. Decision-making approach in decoding nodes

Many implementations of network coding in wireless networks encourage network nodes to listen to the traffic of their neighbors in order to increase coding opportunities in the network. As a part of this strategy, nodes store these packets in their memory for a certain period and then periodically send out reception reports to their neighboring nodes, containing a list of all packets present in its memory. This allows the neighboring nodes to determine the best coding patterns by examining the coding conditions between different packets. An important observation in this context, as highlighted in [12], is that a significant number of these overheard packets do not contribute to improving the coding efficiency in the network.

In this approach, network nodes strive to opportunistically overhear packets that are useful for coding operation over time, rather than indiscriminately capturing all packets. We have formulated this problem as a semi-Markov decision process and the final goal of this problem is to discover a strategy for network nodes to choose between sleep and staying awake in idle states, ultimately maximizing the cumulative reward over time. In brief, in the proposed solution, network nodes learn from their own experiences to determine when to sleep and when to remain awake for the purpose of listening to neighboring packets. Unlike the previous version of this solution, in the upcoming approach, the value of packets is determined not only based on their data content but also on the reception reports they carry. This addresses the undesirable impact of this approach on the decision-making approach in the coding nodes.

#### 1) FORMULATION

The modeling details of the proposed solution are as follows.

**Decision Epochs:** In this problem, decision epochs form a sequence of moments as {$d0$, $d1$, $d2$, ..., $dt$, ...}, where at each epoch an action must be selected, potentially leading to a change in the system's state. In this context, decision epochs represent overhearing opportunities, assumed to follow a random sequence over time. During each overhearing opportunity, a node can decide to remain awake or enter a sleep mode.

**State Space:** The set of possible states for a node is represented as $S$. Additionally, $s_t \in S$, denoting the state of the node at time $t$, is defined as follows:

$$s_t = (e_t, g_t) \quad (15)$$

where $e_t$ represents the remaining energy of the node at time $t$, and $g_t$ signifies the efficiency of network coding operations within its vicinity at time $t$. During implementation, both parameters $e_t$ and $g_t$ need to be quantized into discrete values.

**Action Set:** At each decision epoch, every node decides whether to sleep or stay awake based on the system's state. Here, $A(s_t)$ represents the set of all feasible actions in state $s_t$, and $a_t$ denotes the action taken at time $t$. Each action encompasses the following options:

$$a_t = \begin{cases} 0, & overhear \\ 1, & sleep \end{cases} \quad (16)$$

**Dynamics of States:** Essentially, the dynamics of states can be characterized by two parameters: transition probabilities ($P_{ss'}(a)$) and the expected dwell time in each state for a given chosen action ($F_s$(a)). $P_{ss'}(a)$ signifies that if the system is in state $s$ at the current moment, then after selecting action $a$, with what probability it will transition to state $s'$. Conversely, $F_s(a)$ signifies that if the system is in state $s$ at the current moment and action $a$ is chosen, how long it will take until the next decision epoch occurs. To estimate $F$ in this problem, $Y_{dd'}$ is employed, representing the expected duration between two consecutive decision epochs $d$ and $d'$. It is calculated as follows:

$$Y_{dd'} = \left[ \sum_{i \in K_n} \sum_{j \in (K_i - n)} \lambda_{ij} \right]^{-1} \quad (17)$$

in which $K_n$ represents the set of neighbors of node $n$, and $\lambda_{ij}$ signifies the packet transmission rate from node $i$ to node $j$.

**Optimal Policy:** The policy $\pi = \{(s, a) \mid a \in A, s \in S\}$ constitutes a set of state-action pairs for all states in the semi-Markov decision process. In this context, an optimal policy is one that maximizes the expected cumulative reward.

**Reward Function:** In this solution, a node's reward is directly linked to the energy conservation achieved through a decrease in transmission count, thereby being calculated from the savings in energy consumption.

In wireless nodes, the radio units often constitute the primary source of energy consumption. We have assumed that the energy required by a node to transmit a data packet in one time slot ($T_{slot}$) is given by:

$$E_T = W_{tr} \times T_{slot} \quad (18)$$

where $W_{tr}$ is a constant representing the amount of power each node consumes during transmission. On the other hand, the energy consumption of a node to receive a packet within a duration of $T_{slot}$ is given by:

$$E_R = W_{rc} \times T_{slot} \quad (19)$$

where $W_{rc}$ is a constant denoting the amount power a node consumes for both receiving and overhearing. On the other hand, the energy consumption of a node in the sleep state is considered negligible. With this introduction, if a node chooses to sleep instead of overhearing a packet at a decision epoch, it can save energy consumption by an amount of $E_R$. As mentioned in Section 3, this research assumes that nodes gather energy from the environment through a stochastic process. The amount of energy each node harvests from the environment is at an average rate of $\zeta$ (refer to Fig. 2).

With this background, the amount of reward received by a node at time $t$ is denoted as $r$ and is calculated as follows:

$$r(s_t, a_t) = Reward(s_t, a_t) - Cost(s_t, a_t) \quad (20)$$

where *Cost* and *Reward* are, respectively, equal to:

$$Cost(s_t, a_t) = \begin{cases} 0, & a_t = 1 \\ E_{rc}, & a_t = 0 \end{cases} \quad (21)$$

and

$$Reward(s_t, a_t) = \begin{cases} 0, & a_t = 1 \\ \psi E_{tR}, & a_t = 0 \text{ and } p_t^o \text{ is useful, type one} \\ \xi E_{tR}, & a_t = 0 \text{ and } p_t^o \text{ is useful, type two} \\ 0, & a_t = 0 \text{ and } p_t^o \text{ is useless} \end{cases} \quad (22)$$

In (18), $\psi$ $(0 \le \psi)$ represents the utility factor for the first type of usefulness, and $\xi$ $(0 \le \xi)$ stands for the utility factor for the second type of usefulness for the overheard packets. Here, the usefulness of overheard packets is considered in two scenarios. In the first type of usefulness, a node overhears a beneficial packet and informs neighboring nodes about receiving it by broadcasting a report. This packet contributes to the coding operations of neighboring nodes (similar to packet $a$ being overheard by node $n2$ at time $t_i$ in Fig. 1). However, in the second type of usefulness, a node overhears a packet containing valuable reception report, which assists the node in discovering new coding opportunities. For instance, packet $r$-$r$, containing useful reception report, is overheard by node $R$ at time $t_k$ in Fig. 1.

The factor $\psi$ corresponds to the number of times an overheard packet participates in the coding operations of neighboring nodes. In this case, the first type of usefulness for the packets is defined as follows:

$$p_t^o = \begin{cases} the\ first\ type\ of\ useful, & if\ p_t^o\ is\ in\ p_k^c \exists p_k^c,\ t \le k \le t + \alpha \\ other\ types\ of\ packet, & else \end{cases} \quad (23)$$

where $p_k^c$ represents the received coded packet at time $k$, and $p_t^o$ is the packet overheard at time $t$. Additionally, $\alpha$ denotes the time a node retains an overheard packet in memory, during which the usefulness of the packet is computed (counted). In a broader context, when a coded packet has participated in coding operations for $\psi$ coded packets, it has reduced the number of transmissions in the network by $\psi$ counts.

Furthermore, parameter $\xi$ for a reception report is equal to the number of useful packets present for coding within that report. This definition suggests that when an overheard packet includes a reception report confirming the receipt of several packets of the first type of usefulness from a neighboring source, it is categorized as the second type of useful packet. The count of these reported packets is determined by the parameter $\xi$. In practice, this number $\xi$ is added to the coding opportunities in a coding-node.

#### 2) REINFORCEMENT LEARNING SOLUTION

There are numerous algorithms for solving semi-Markov Decision Process, aiming to find optimal solutions [41]. Among these, Reinforcement Learning (RL) provides a viable approach for addressing decision-making problems that are theoretically challenging to find optimal solutions for. Among the reinforcement learning methods, the Q-learning technique is the most recognized and widely used method [42]. In order to solve our current problem, we have introduced a modified version of continuous-time Q-learning [43].

It's worth noting that in our problem, when a node makes the decision to overhear in a decision epoch, the reward associated with this action is not immediately apparent. The node must wait until the usefulness of the overheard packet is determined to ascertain the reward of its action. We have modified the continuous-time Q-learning algorithm in a way that addresses the issue of delayed rewards as well.

In our continuous-time Q-learning, the Q-values at decision epoch $d_{t+1}$ are updated using the following expression:

$$Q(s_t,a_t) = Q(s_t,a_t) + \beta \times \left(R(s_{t+1},a_{t+1}) + e^{-\gamma(d_{t+1}-d_t)} \max_{a_{t+1}\in A(s_{t+1})} \{Q(s_{t+1},a_{t+1})\} - Q(s_t,a_t)\right) \quad (24)$$

In our problem, when a node selects action $a_t = 0$ at decision epoch $d_t$ and overhead a packet, the reward associated with this packet is unknown. The node has to wait for the impact of this packet and the level of its usefulness for coding operations to become clear. The potential reward is proportional to the number of contributions of the overheard packet in the coding operation. These rewards might even be received out of sequence.

In this learning approach, to overcome the challenge of delayed rewards, the time delay dimension needs to be incorporated into the learning process. Therefore, it is assumed that after a node selects an action, the delay until receiving the reward for that action involves several decision epochs, where this count follows a Poisson random variable. Consequently, a dimension for delay needs to be added to the Q-values. Each entry in the Q-table is represented as $Q(s, a, \hat{\theta})$, where $\hat{\theta}$ is a discrete-valued variable representing the delay in terms of the number of decision epochs. Essentially, in this method, the node needs to retain a suitable number of performed actions in its history to preserve delayed rewards for a certain duration.

In practice, a node can maintain multiple Q-tables for different delay values. Throughout the learning process, a node aims to learn about the delay of rewards and then uses the estimated delay value during decision epochs.

The node selects the most valuable estimate of the delay ($\hat{\theta}^*$) as follows:

$$\hat{\theta}^* = \arg\max_{\hat{\theta}} Q(s,a,\hat{\theta}) \quad (25)$$

and after that, assuming that the precise delay value is exactly equal to this estimate, the node should consider all Q-value entries that have a delay equal to $\hat{\theta}^*$. Among these entries, it chooses the most valuable action to execute as follows:

$$a^* = \arg\max_{a} Q(s,a,\hat{\theta}^*) \quad (26)$$

After selecting and executing an action, the system transitions to state $s'$, and the entire process is repeated from scratch.

In this manner, during the course of learning, a node doesn't have knowledge of the actual delay of rewards and must assign appropriate credits to all Q-values. To achieve this, the node updates the Q-values at each iteration and allocates the corresponding rewards to all of them. To update the Q-values, we utilize the following relationship:

$$Q(s_{t-\hat{\theta}}, a_{t-\hat{\theta}}, \hat{\theta}) \xleftarrow{update} Q(s_{t-\hat{\theta}}, a_{t-\hat{\theta}}, \hat{\theta}) + \beta \times \Big(R(s_t,a_t) + e^{-\gamma(d_{(t+1)-\hat{\theta}}-d_{t-\hat{\theta}})} \max_{a_{t+1}\in A(s_{t+1}),\hat{\theta}} \{Q(s_{(t+1)-\hat{\theta}}, a_{(t+1)-\hat{\theta}}, \hat{\theta})\} - Q(s_{t-\hat{\theta}}, a_{t-\hat{\theta}})\Big) \quad (27)$$

wherein $\hat{\theta}$ values are selected in order from among all estimates. A summary of the notation in this research is provided in Table I.

### C. Required changes in IEEE 802.11

As previously mentioned, in the proposed framework, nodes utilize the IEEE 802.11 unicast mode [44]. However, some modifications and enhancements need to be implemented in the unicast mode to ensure its compatibility with the proposed framework. The following points will cover these changes and considerations:

**First change:** When sending a coded packet, the destination field in the data link layer will be filled with one of the destinations of the coded packets, and the remaining destinations of the pure packets will be added to the coding header. When a node (in a promiscuous mode) receives a coded-packet with a destination address different from its own MAC address, it needs to examine the coding header to determine if it is listed as a subsequent coding step or not. If it finds its address in the coding header and in the list of subsequent hops, it starts processing the packet as a received coded packet. Consequently, only the neighbor designated as the packet's destination returns the acknowledgment of receiving the packet. In this scenario, if confirmation of receipt is not received from the primary receiver within a specified time interval, the packet is retransmitted. With this method, collision occurrence in the primary receiver is detected and mitigated, but the collision (or non-reception) of this packet in other nodes is disregarded. For this reason, we require the following two enhancements.

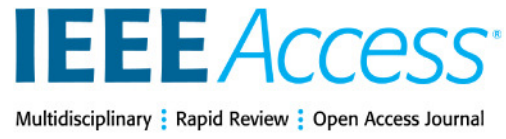

TABLE I
A SHORT SUMMARY OF NOTATION

| Symb. | Meaning | Symb. | Meaning |
|---|---|---|---|
| $D$ | Current best coding degree | $s_t$ | System state in time $t$ |
| $d^*$ | Optimal coding degree to stop | $a_t$ | Taken action in time $t$ |
| $\lambda_r$ | Arrival rate of reception report | $d_t$ | The best coding degree in $t$ |
| $Y_{dd'}$ | Time between $d$ and $d'$ | $T_{slot}$ | Duration of transmission |
| $F_s(a)$ | Sojourn time in $s$ by action $a$ | $W_{rc}$ | Receiving power |
| $p_{i,j}$ | Transition prob. From $s$ to $s'$ | $W_{tr}$ | Transmission power |
| $r()$ | Reward function | $\hat{\theta}$ | Estimate of delay |
| $p_t^o$ | Overheard packet in time $t$ | $\hat{\theta}^*$ | Most valuable delay estimate |
| $p_t^c$ | Received coded-packet in time $t$ | $E_I$ | Idle listening energy consumption |
| $\psi$ | First type of usefulness factor | $E_R$ | Receiving energy consumption |
| $\xi$ | Second type of usefulness factor | $E_T$ | Sending energy consumption |

**Second change:** To enhance the confidence of receiving the coded-packet by the remaining destinations, we leverage the RTS/CTS mechanism. In general, these messages are used to address the hidden terminal problem and the exposed terminal problem [45]. Before sending a data packet, the source sends an RTS packet containing the destination and the size of the original data packet. The destination responds with a CTS packet, and then the source begins transmitting the actual data packet upon receiving the CTS packet and waits for acknowledgment from the destination. The other nodes that hear the exchange of RTS and CTS packets are required to postpone their data transmission out of respect for this pair. It's worth noting that the RTS/CTS exchange significantly reduces the likelihood of collisions at the destination, even when collisions in other listening nodes aren't detectable. However, employing this technique has an additional beneficial side effect: a packet may be resent multiple times until it's received by the primary destination, which increases the chances of other destinations overhearing these packets.

**Third change:** In contrast to the unicast mode in IEEE 802.11 where only the primary destination of a packet acknowledges its reception, here, when a coded-packet is sent, its reception needs to be acknowledged by all nodes interested in receiving that coded packet (which require one of the pure packets within it). To achieve this, we employ local retransmissions. In this method, the sender expects all recipients of the coded-packet (who need one of the pure packets within it) to acknowledge its reception. If one or more of these recipients do not acknowledge the reception of this packet within a specified time interval, the coded-packet will be resent. This resent packet might even be a new combination of coding, and there's no longer a need for the participation of the pure packets whose recipients acknowledged the reception of the previous coded-packet.

**Fourth change:** In IEEE 802.11's unicast mode, for sending acknowledgment packets, a synchronous method is used, meaning that as soon as a node receives a packet, it must inform the source node by sending a new acknowledgment packet. Unfortunately, in practice, sending acknowledgments like this for coded-packets incurs significant costs and overhead due to the potentially large number of acknowledgment packets. To address this, we use asynchronous piggybacking method for sending acknowledgments. In this approach, the receiving node doesn't send separate acknowledgment packets immediately towards the source; instead, it includes acknowledgments along with its data packets. This way, the acknowledgment overhead is reduced since acknowledgments are combined with regular data transmission.

## VI. Numerical Results

In this section, we have compared the performance of the ENCODE framework with two other approaches, COPE and a non-coding approach. First, we explain the simulation environment, and then we present and analyze the results obtained from the simulation process.

### A. Simulation environment

Simulation was carried out using NS2 [46]. A total of 200 stationary nodes were randomly deployed in an environment with dimensions of 1500 meters by 1500 meters. The signal attenuation of nodes follows the Two-way ground modeling, and each node has a nominal transmission range of 225 meters. The network nodes operate in the MAC layer using a modified IEEE 802.11 protocol. For the routing algorithm, a simple geometric algorithm is utilized. To generate traffic, two nodes are randomly selected as the source and destination of a flow. Then, a UDP flow is established between these two nodes using packets of a fixed size (450 bytes), and the inter-packet time interval follows an exponential distribution.

In this simulation, for the implementation of the first decision-making approach, the parameter $L$ was set to 22 packets and the delay discount factor ($\delta$) was set to 0.05. For implementing the second approach of decision-making, the learning rate ($\beta$) was set to 0.45, the discount factor ($\gamma$) was set to 0.85, and the maximum acceptable delay for rewards based on decision epochs ($\theta_{max}$) was set to 8. To reduce the state space, the parameters $g_t$ and $e_t$ were quantized as follows: the parameter $e_t$ was quantized to integer values between 1 and 8, and the parameter $g_t$ was considered between 1 and 10. To estimate the parameter $\lambda_d$, an adaptive LMS filter with a parameter of 4 was used.

### B. Simulation Results

#### 1) CONVERGENCE OF THE RL SOLUTION

The convergence time of the presented reinforcement learning algorithm in Section 5-2 has been evaluated and illustrated in Fig. 3 based on the average rewards obtained. As observed in the plot, this algorithm requires approximately 55,000 iterations to complete the learning process and achieve an optimal policy. It's important to note that the convergence time of reinforcement learning algorithms depends on various parameters during runtime, such as learning factor, size of state space, size of action space, and so on. In this simulation, the average network traffic volume has been set to 800 kbps.

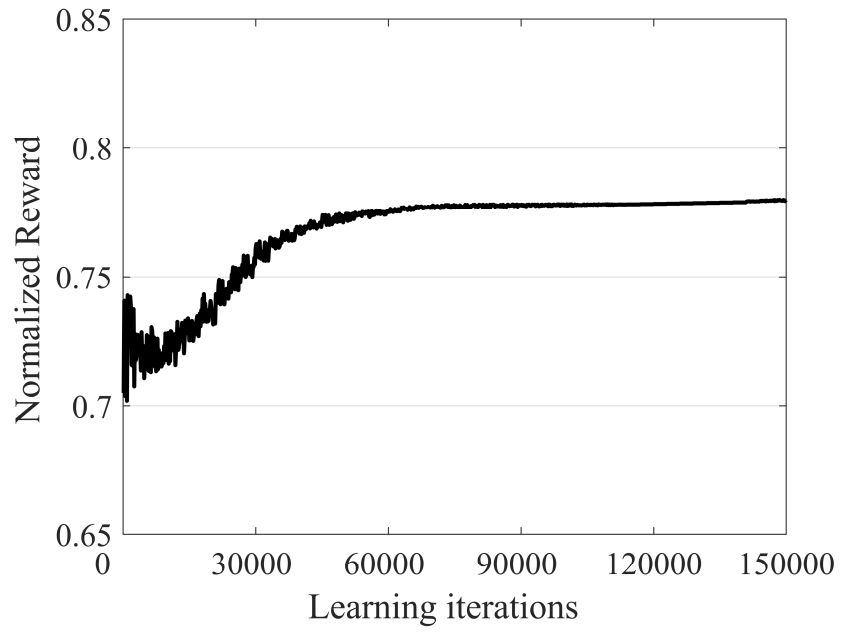

**FIGURE 3.** Learning approach convergence

## 2) AVERAGE CODING GAIN

One important aspect of evaluating the proposed framework against existing approaches is to compare them in terms of average coding gain. In Fig. 4, the coding gain of ENCODE and COPE is compared against the absence of network coding. In both ENCODE and COPE approaches, as the number of flows and packets in the network increases, the coding opportunities also increase, leading to more packet combinations. In the ENCODE framework, with the increase in the number of flows in the network, the average coding gain becomes slightly higher than COPE, based on the coding-node's strategy to find better coding opportunities.

In Fig. 5, the impact of node density on coding gain in the network is depicted. In this scenario, the average traffic rate generated in the network is set to 1000 Kbps. In this case, as the node density increases, more coding opportunities arise for intermediate nodes, leading to an improvement in the coding gain.

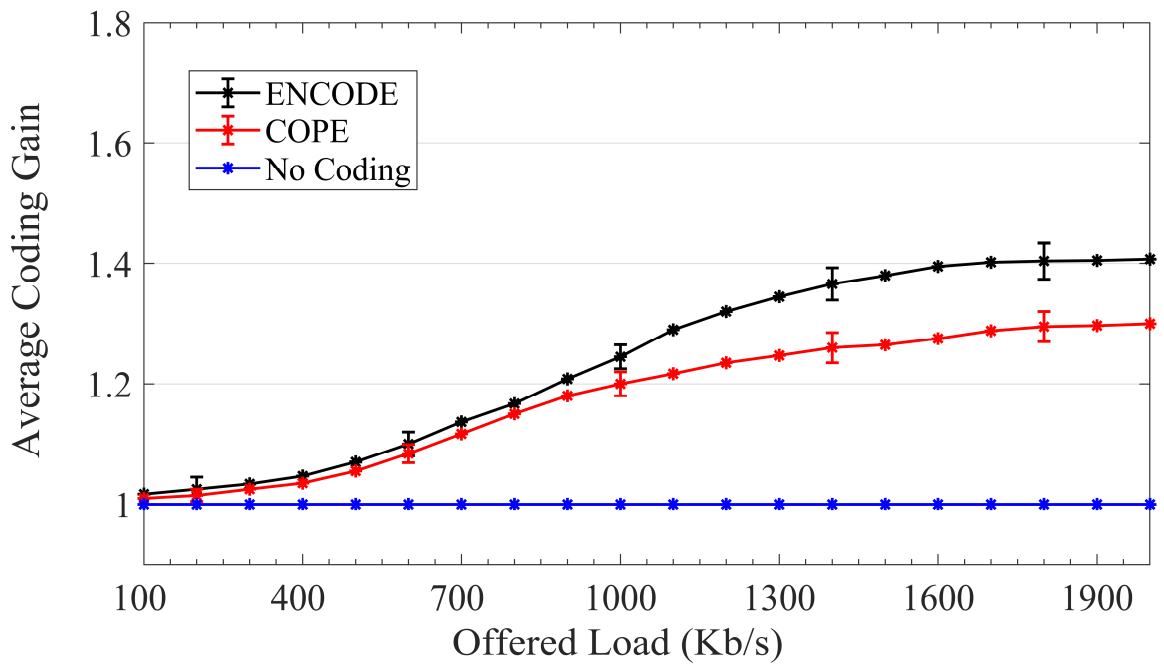

**FIGURE 4.** Coding gain vs. traffic load.

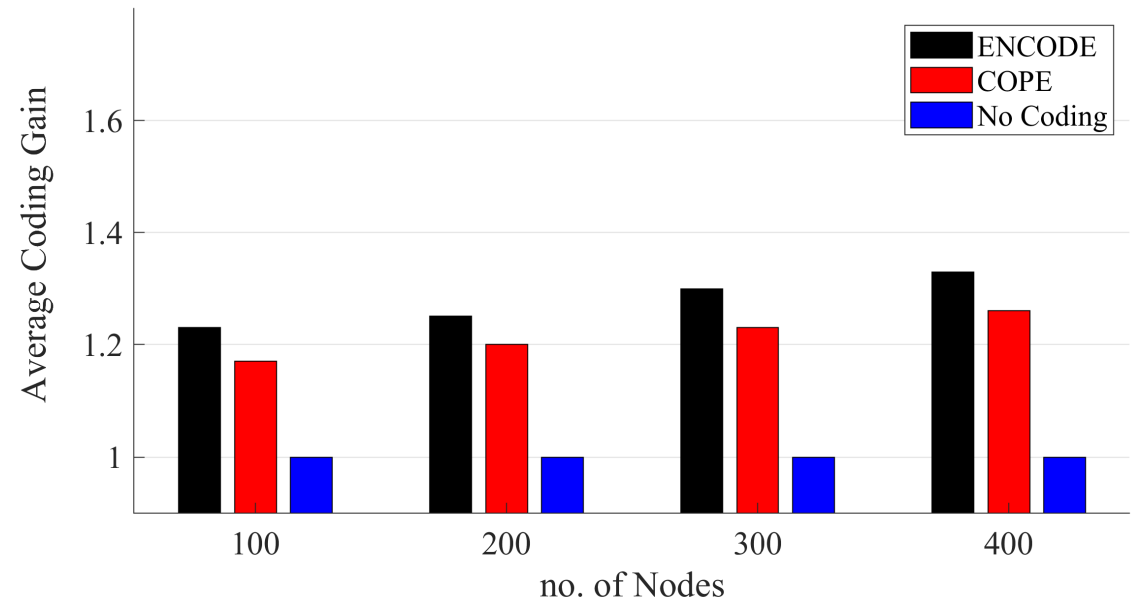

**FIGURE 5.** Coding gain vs. node density.

## 3) NETWORK THROUGHPUT

In this context, throughput is defined as the rate of successful packet delivery to the destination. When increasing the number of flows and consequently the network traffic volume without using coding, the network approaches its saturation state, and as traffic volume continues to increase, the throughput decreases. This effect occurs with a slight delay when using network coding, and the network takes longer to approach its saturation state. In the comparison between ENCODE and COPE, the proposed approach demonstrates better throughput due to its higher coding gain. The comparison of the three discussed approaches in terms of network throughput is depicted in Fig. 6.

In Fig. 7, the impact of node density on network throughput is illustrated. For this purpose, the input traffic rate to the network has been set to 1000 Kbps. Higher node density means more potential paths and more nodes through which data can be encoded. For this reason, with the increase in node density, the network throughput slightly increases.

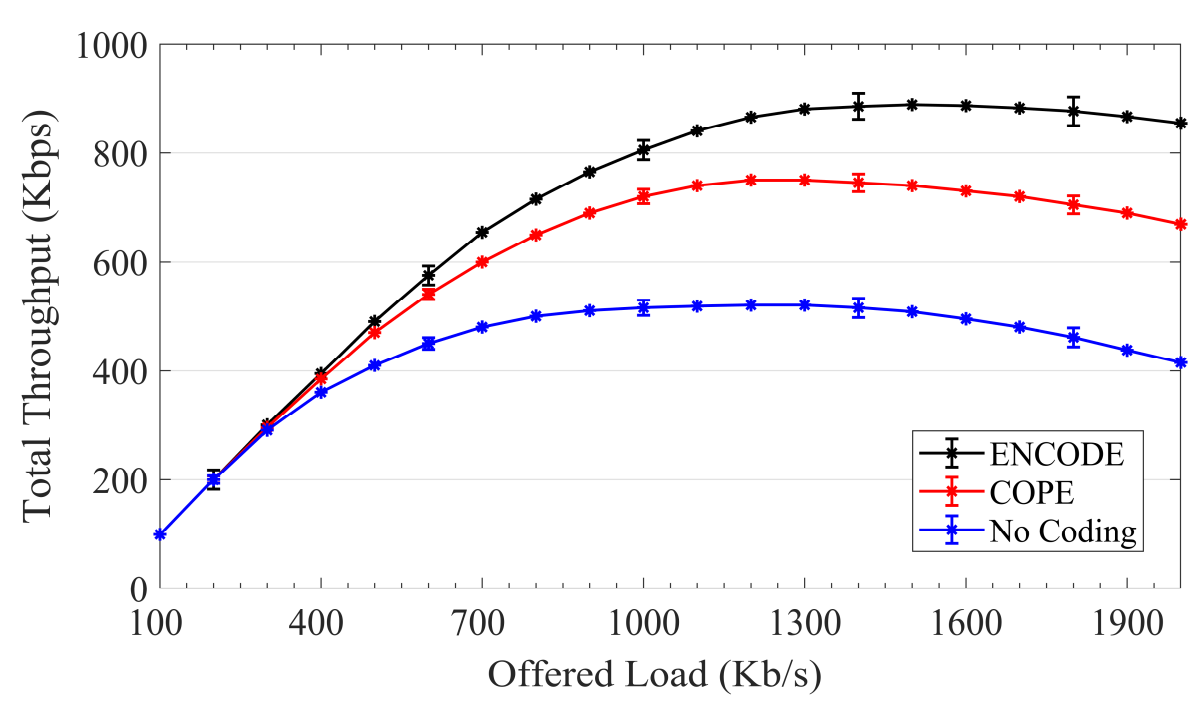

**FIGURE 6.** Network throughput vs. traffic load.

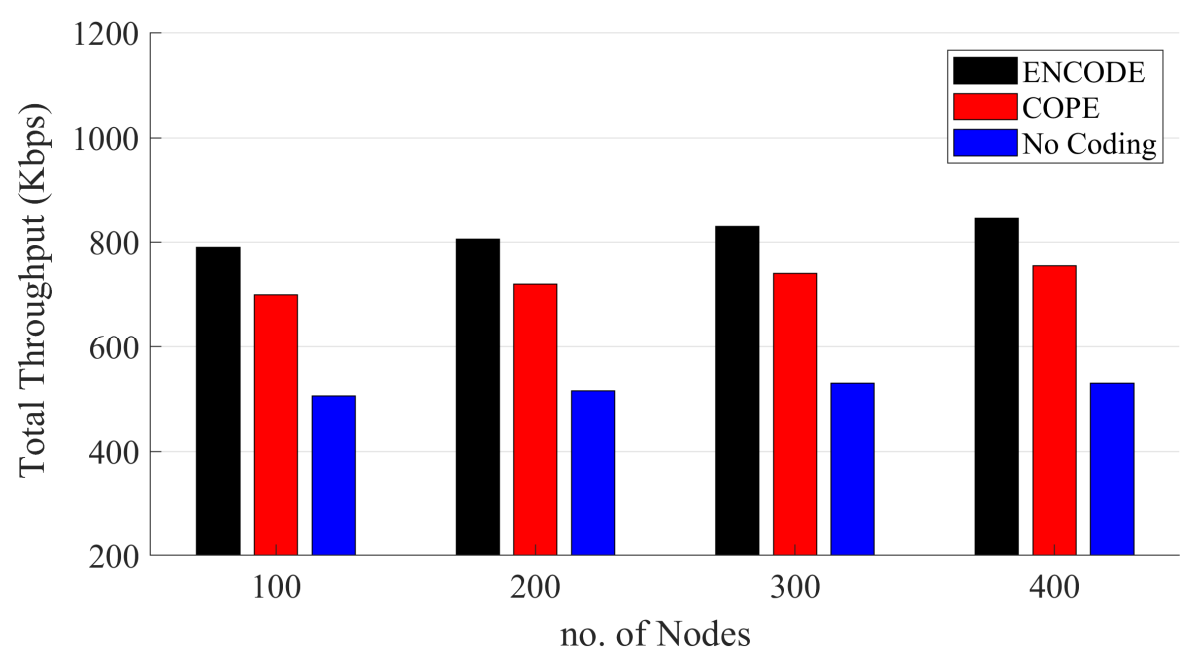

**FIGURE 7.** Network throughput vs. node density.

## 4) AVERAGE END-TO-END DELAY

Here, the total delay of packet transmission from the moment of sending at the source to the moment of reception at the destination is referred to as end-to-end delay. When the network traffic is low, collisions are negligible, and therefore the packet delays in the three examined methods are similar, especially since coding opportunities rarely arise in low traffic scenarios. However, as the network traffic volume increases, on one hand, the probability of collisions rises, and on the other hand, more coding opportunities arise in the network. In the ENCODE framework, due to the operation of coding-nodes, a slight delay is added to the packet transmission time. Fortunately, thanks to the higher coding gain and better packet combining, leading to a reduction in the number of transmitted packets, a portion of this delay is effectively compensated. Fig. 8 provides a comparison of the three mentioned approaches in terms of end-to-end delay.

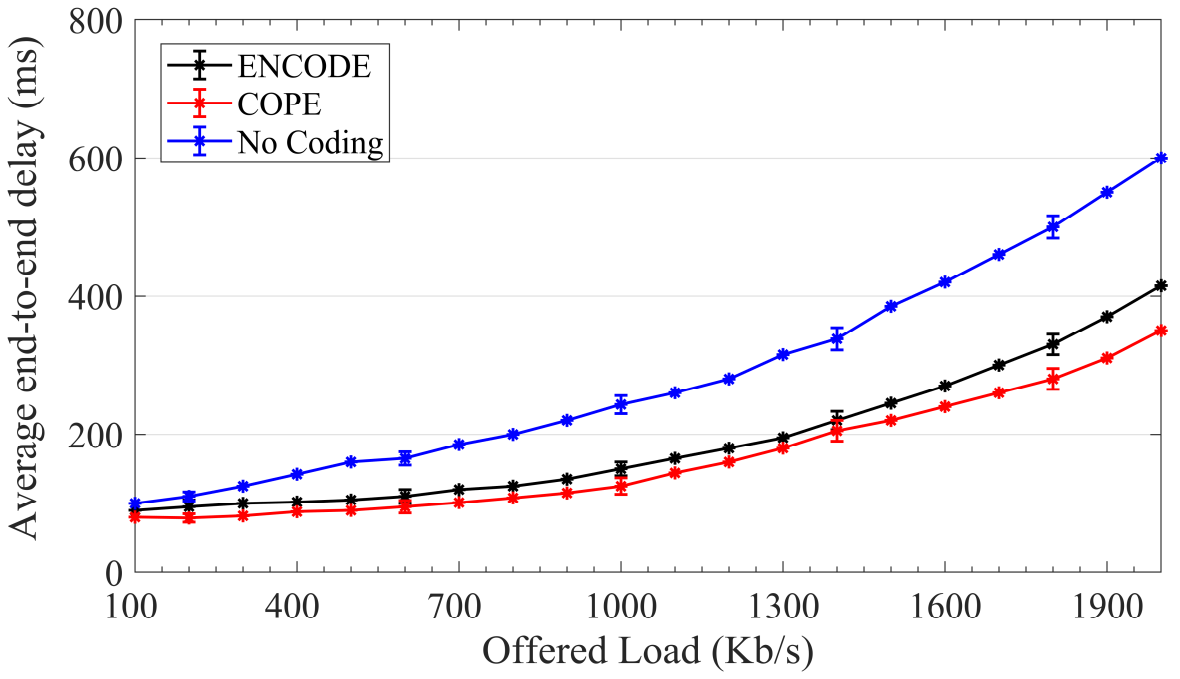

**FIGURE 8.** **End-to-end delay vs. traffic load.**

### 5) ENERGY CONSUMPTION

In this study, the average energy consumption per bit has been used as the energy consumption metric for comparing the three approaches. Due to various overheads, as the network traffic volume increases, the average energy consumption per bit delivered to the destination gradually increases throughout the network in all approaches. This increase in energy consumption becomes more pronounced at higher network traffic volumes due to increased collisions and retransmissions. In the ENCODE method, due to the overhearing approach of the nodes, efforts are made to conserve energy, which can be clearly observed in comparison with the COPE method in Fig. 9.

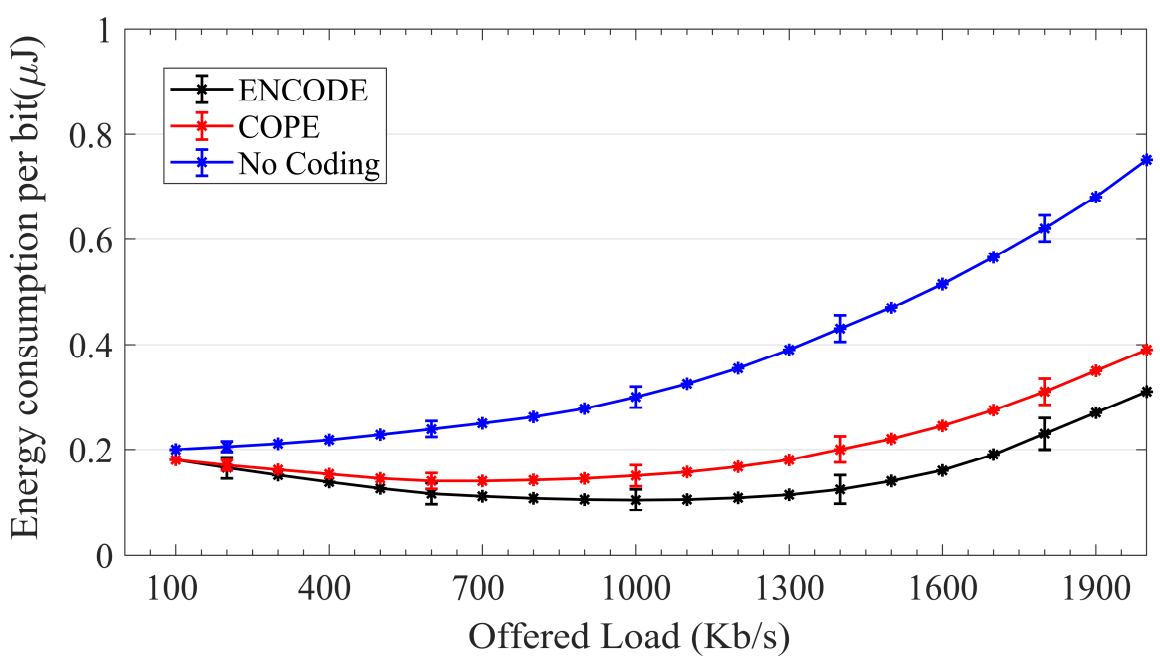

**FIGURE 9.** **Energy consumption vs. traffic load.**

## VII. Conclusion

The mission of this paper was to provide an efficient integrated framework for utilizing network coding in multi-hop wireless networks, based on the simultaneous enhancement of two decision-making approaches in coding nodes and decoding nodes. To achieve this, the system model was redefined to precisely evaluate the impact of the two decision-making approaches on each other's performance. In the proposed framework, named ENCODE, coding nodes determine the optimal transmission time for coded packets using optimal stopping theory, striking a trade-off between delay and coding gain. On the other hand, decoding nodes learn over time using SMDP and a reinforcement learning-based solution to identify which packets are worth opportunistically overhearing, considering a trade-off between energy consumption and coding gain. Finally, the proposed framework was compared with the COPE framework (as a reference model) through simulation.

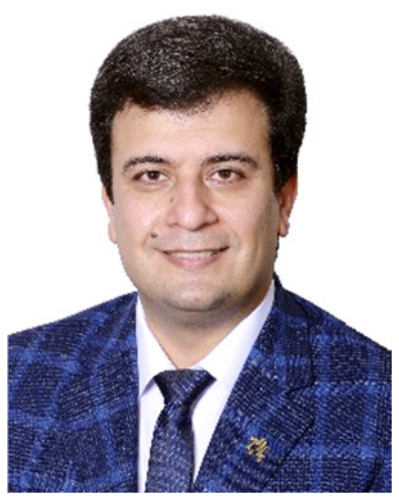

**Nastooh Taheri Javan** (*Senior Member, IEEE*) is an Assistant Professor with the Computer Engineering department at Imam Khomeini International University (IKIU), Qazvin, IRAN. Dr. Taheri Javan was post-doctoral fellow at Amirkabir University of Technology (Tehran Polytechnic), Tehran, IRAN, where he completed his M.S. and Ph.D. in computer engineering in 2007 and 2017, respectively. His research interests include wireless networks, network coding and optimization.

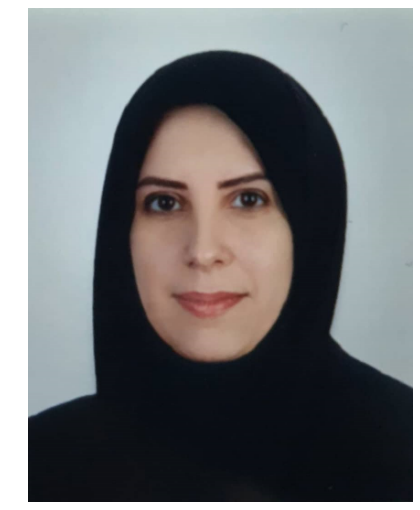

**Zahra Yaghoubi** received the B.S. and M.S. degrees in Electrical Engineering from Imam Khomeini International University, Qazvin, Iran and Ph.D. in Electrical Engineering from Amirkabir University of Technology, Tehran, Iran. Now she is an assistant professor at Imam Khomeini International university, Qazvin, Iran. Her research interest includes nonlinear control, multi-agent control, robotic and adaptive control.